%% file: P040050-06.tex
\documentclass[prd,twocolumn,eqsecnum,superscriptaddress,showpacs,letterpaper,preprintnumbers,altaffilletter]{revtex4}
\usepackage{amsmath}
\usepackage{bbold}
\usepackage{color}
\usepackage{fancyhdr}
\usepackage{float}
\usepackage[dvips]{graphicx}
\usepackage{latexsym}
\usepackage{times}

\newcommand{\be}{\begin{equation}}
\newcommand{\ee}{\end{equation}}

\newcommand{\bea}{\begin{eqnarray}}
\newcommand{\eea}{\end{eqnarray}}
\newcommand{\bean}{\begin{eqnarray*}}
\newcommand{\eean}{\end{eqnarray*}}
\newcommand{\bi}{\begin{itemize}}
\newcommand{\ei}{\end{itemize}}

\newcommand{\ind}[1]{\mbox{\scriptsize{#1}}}

\begin{document}

\title{Upper limits from the LIGO and TAMA detectors on the
rate of gravitational-wave bursts}

\input{ligotama_auth_v5}

\date{\today}

\begin{abstract}

We report on the first joint search for gravitational waves by the
TAMA and LIGO collaborations.  We looked for millisecond-duration
unmodelled gravitational-wave bursts in 473 hr of coincident data
collected during early 2003.
No candidate signals were found.  We set an upper limit of 0.12
events per day on the rate of detectable gravitational-wave bursts, 
at 90\% confidence level.  From simulations, we estimate that our
detector network was sensitive to bursts with root-sum-square strain 
amplitude above approximately  $1$-$3\times10^{-19}$ Hz$^{-1/2}$
in the frequency band 700-2000 Hz.
We describe the details of this collaborative search, with particular
emphasis on its advantages and disadvantages compared to searches by
LIGO and TAMA separately using the same data.  
Benefits include a lower background and longer observation time, at
some cost in sensitivity and bandwidth.
We also demonstrate techniques for performing coincidence searches
with a heterogeneous network of detectors
with different noise spectra and orientations.  
These techniques include using coordinated signal injections to estimate 
the network sensitivity, and tuning the analysis to maximize 
the sensitivity and the livetime, subject to constraints on the
background.
 
\end{abstract}

\preprint{LIGO-P040050-06-Z}

\maketitle


\section{Introduction}
\label{sec:introduction}

At present several large-scale interferometric gravitational-wave
detectors are operating or are being commissioned: GEO
\cite{Wi_etal:04}, LIGO \cite{Si:04}, TAMA \cite{Ta:04}, and Virgo
\cite{Ac_etal:04}.  In addition, numerous resonant-mass detectors have
been operating  for a number of years
\cite{Am_etal:89,Al_etal:00,As_etal:03}.  Cooperative analyses by
these observatories could be valuable for making confident detections
of gravitational waves and for extracting maximal information from
them.  This is particularly true for gravitational-wave bursts (GWBs)
from systems such as core-collapse supernovae
\cite{ZwMu:97,DiFoMu:02a,DiFoMu:02b,OtBuLiWa:04}, black-hole mergers
\cite{FlHu:98a,FlHu:98b}, and gamma-ray bursters \cite{Me:02}, for
which we have limited theoretical knowledge of the source and 
the resulting gravitational waveform to guide us.  Advantages
of coincident observations include a decreased background 
from random detector noise fluctuations, an increase in the total observation
time during which some minimum number of detectors are operating, and
the possibility of locating a source on the sky and extracting polarization
information (when detectors at three or more sites
observe a signal) \cite{GuTi:89}.
Independent observations using different detector hardware and
software also decrease the possibility of error or bias.

There are also disadvantages to joint searches.  Most notably, in a
straightforward coincidence analysis the sensitivity of a network is
limited by the least sensitive detector.  In addition, differences in
alignment mean that different detectors will be sensitive to different
combinations of the two polarization components of a  gravitational
wave.  This complicates attempts to compare the signal amplitude or
waveform as measured by different detectors.
Finally, differences in hardware, software, and algorithms make
collaborative analyses technically challenging.

In this article we present the first observational results from a 
joint search for gravitational
waves by the LIGO and TAMA collaborations.  We perform a coincidence 
analysis targeting generic millisecond-duration GWBs, requiring 
candidate GWBs to be detected by all operating LIGO and TAMA
interferometers. 
This effort is complementary to searches for GWBs
performed independently by LIGO \cite{Ab_etal:05} and TAMA
\cite{An_etal:04} using the same data that we analyze here.
Our goal is to highlight the strengths and weaknesses of 
our joint search relative to these single-collaboration searches, 
and to demonstrate techniques for performing coincidence searches
with a heterogeneous network of detectors
with different noise spectra and orientations.
This search could form a prototype for more comprehensive
collaborative analyses in the future.

In Section~\ref{sec:data} we review the performance of the LIGO and 
TAMA detectors during the joint observations used for this search.
We describe the analysis procedure in Section~\ref{sec:method}, and 
the tuning of the analysis in Section~\ref{sec:sims_tuning}.
The results of the search are presented in Section~\ref{sec:results}. 
We conclude with some brief comments in Section~\ref{sec:summary}.


\section{LIGO-TAMA Network and Data Sets}
\label{sec:data}

The LIGO network consists of a 4 km interferometer ``L1'' near 
Livingston, Louisiana and 4 km ``H1'' and 2 km ``H2'' interferometers which
share a common vacuum system on the Hanford site in Washington.  The TAMA group operates
a 300 m interferometer ``T1'' near Tokyo.  These instruments attempt to
detect gravitational waves by  monitoring the interference of the
laser light from each of two perpendicular arms.  Minute differential
changes in the arm lengths produced by a  passing gravitational wave
alter this interference pattern.  Basic information on the position
and orientation of the LIGO and TAMA detectors can be found in
\cite{Al_etal:01,AnBrCrFl:01}.  Detailed descriptions of their
operation can be found in \cite{Si:04,Ab_etal:05,An_etal:04,Ab_etal_NIM:04}.

In a search for gravitational-wave bursts, the key characteristics
of a detector are the orientation, the noise spectrum and its
variability, and the observation time.

The response of an interferometer to a gravitational wave depends on 
the relative orientation of the source and the
detector, as well as on the signal polarization.
Figure~\ref{fig:antenna} shows the variation in the polarization-averaged 
sensitivities of the LIGO and TAMA detectors as a function of the sky
position of the source.  It is clear from these figures that LIGO and 
TAMA have maximum sensitivity to different portions of the sky.
This complicates a search based on
coincident detections: there is a loss of sensitivity to weak
signals; and it is difficult to compare quantitatively the signal
amplitude or waveform as measured by the LIGO and TAMA detectors since they
will not, in general, be the same.  
(This was not a significant problem in previous multi-detector searches by 
LIGO \cite{Ab_etal:04,Ab_etal:05} and the IGEC \cite{As_etal:03}, since
they employed approximately co-aligned detectors.)   
We account for these effects by using coordinated simulations
to guide the tuning of our analysis so as to maximize the
detection efficiency of the network, and we forego amplitude
and waveform consistency tests between LIGO and TAMA; see
Sections~\ref{sec:method}, \ref{sec:sims_tuning}.

\setlength{\unitlength}{1in}
\begin{figure*}[tbp] 
  \begin{center}
    \begin{picture}(6,7)
      \put(0,4.8){\rotatebox{0}{\resizebox{6in}{!}{\includegraphics{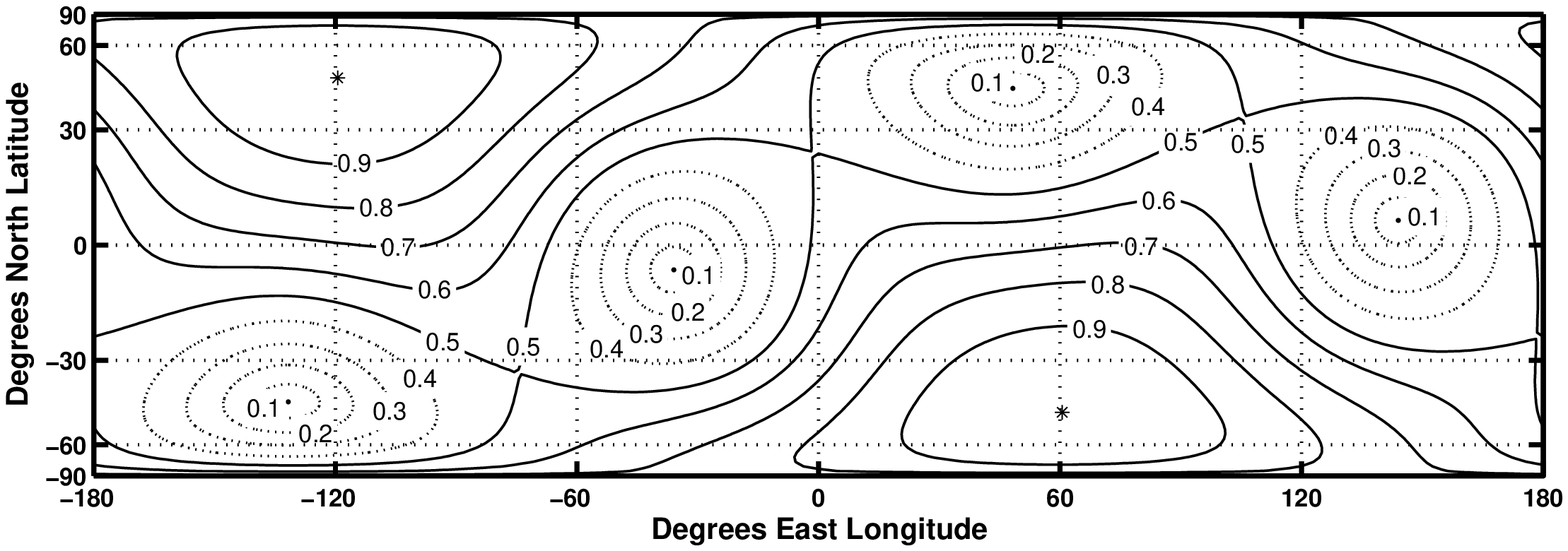}}}} 
      \put(0,2.4){\rotatebox{0}{\resizebox{6in}{!}{\includegraphics{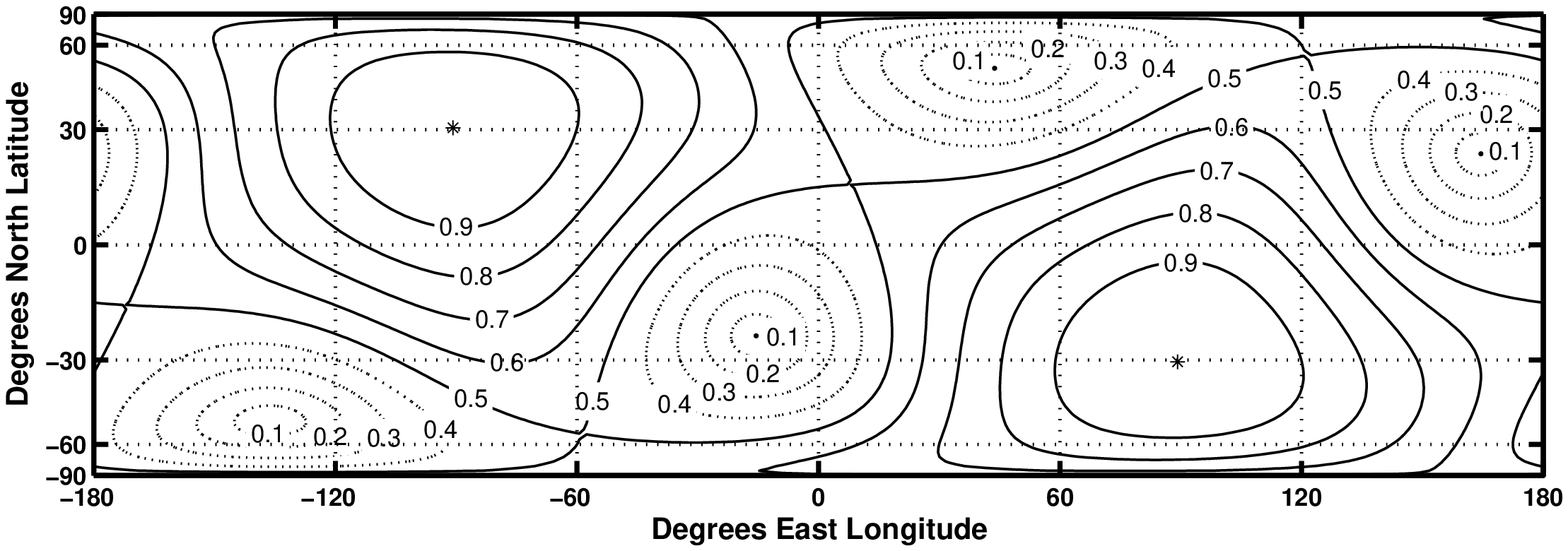}}}} 
      \put(0,0){\rotatebox{0}{\resizebox{6in}{!}{\includegraphics{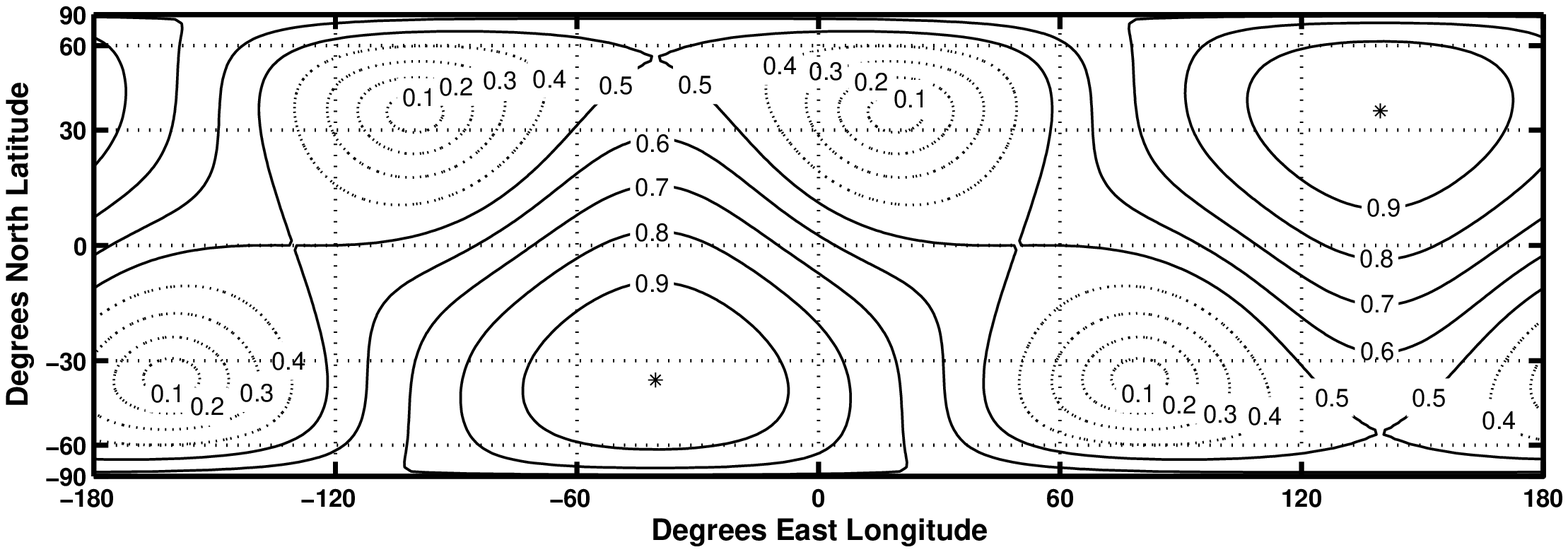}}}} 
    \end{picture}
    \caption{\label{fig:antenna}  
    Polarization-averaged antenna amplitude responses 
    $\left(F_+^2+F_\times^2\right)^{1/2}\in[0,1]$, 
    in Earth-based coordinates.  [See equation (\ref{eqn:h_of_t}) and 
    \cite{AnBrCrFl:01} for definitions of these functions and 
    of Earth-based coordinates.]
    The top plot is for the LIGO Hanford detectors (H1, H2).
    The middle plot is for LIGO Livingston (L1).  The bottom plot is for
    TAMA (T1).   High contour values indicate sky directions of high
    sensitivity.  The directions of maximum (null) sensitivity for each detector 
    are indicated by the * (.) symbols.  
    The directions of LIGO's maximum sensitivity lie
    close to areas of TAMA's worst sensitivity and vice versa.  }
  \end{center}
\end{figure*}

\begin{figure}[tbp] 
  \begin{center}
    \includegraphics[width=8.5cm]{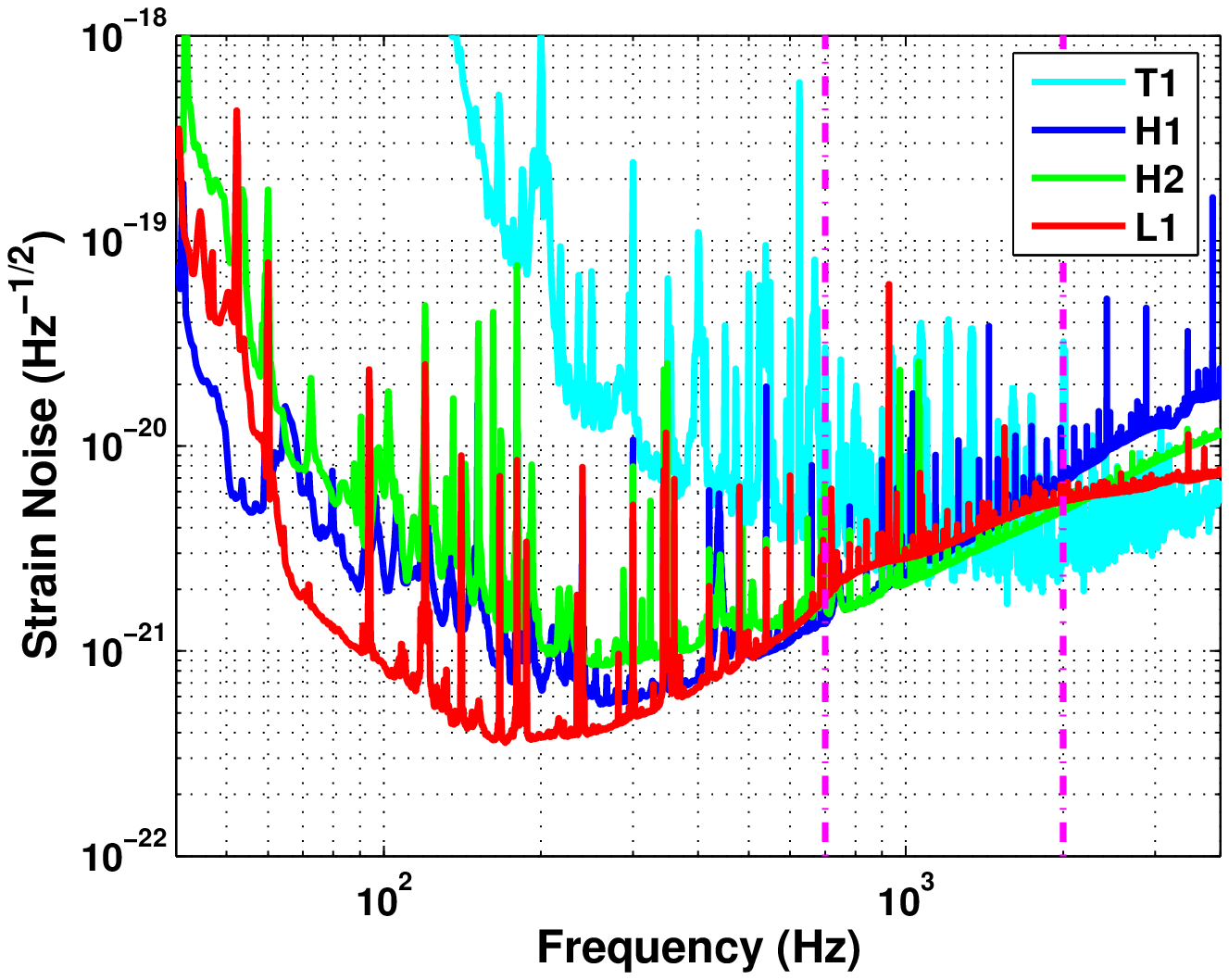}
    \caption{\label{fig:spectra}
      S2-averaged amplitude noise spectra for the LIGO detectors,  
      and a representative DT8 spectrum for the TAMA detector.  
      We focus on GWBs which have significant energy in the frequency range 
      700-2000 Hz (indicated by the vertical dashed lines), where each  
      interferometer has approximately the same noise level.
    }
  \end{center}
\end{figure}

\begin{figure}[tbp] 
  \begin{center}
    \includegraphics[width=8.5cm]{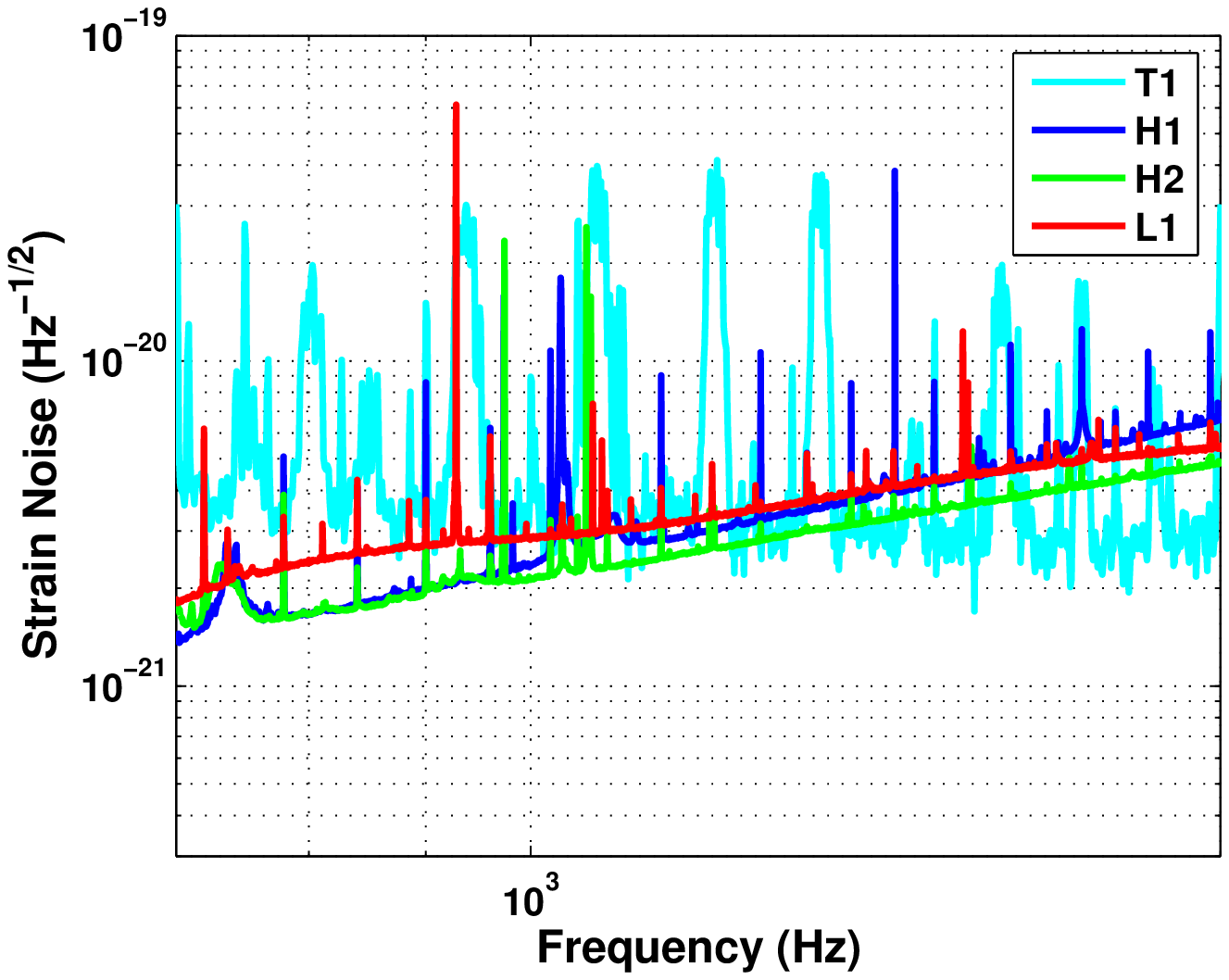}
    \caption{\label{fig:spectra_zoom}
      The same amplitude noise spectra as in Figure~\ref{fig:spectra}, 
      focusing on the frequency range 700-2000 Hz.  
      The peaks at multiples of 400/3 Hz in the TAMA spectrum are due 
      to a coupling between the radio-frequency modulation signal and 
      the laser source; these frequencies are removed   
      by the data conditioning discussed in Section~\ref{sec:power}.
    }
  \end{center}
\end{figure}

The data analyzed in this search were collected during the 
LIGO science run 2 (S2) and the TAMA data taking run 8 (DT8), 
between 14 February 2003 and 14 April 2003.  
Figures \ref{fig:spectra} and \ref{fig:spectra_zoom} show
representative strain noise spectra from each detector during S2/DT8.
Ignoring differences in antenna response, requiring coincident detection of
a candidate signal by both LIGO and TAMA means that the sensitivity
of the network will be limited by the least sensitive detector.  This
motivates concentrating our efforts on the frequency band where all detectors
have comparable sensitivity; i.e., near the minimum of the noise
envelope.  Specifically,  we choose to
search for GWBs that have significant power in
the  frequency range 700-2000 Hz.  Restricting the frequency range in
this manner reduces the background due to coincident noise
fluctuations,  while preserving the sensitivity of the network to
GWBs that are detectable by both LIGO and TAMA.
Note also that the LIGO collaboration has carried out an independent
GWB analysis of the S2 data concentrating on the band 100-1100 Hz
\cite{Ab_etal:05}.  There is thus no danger in missing a real
detectable burst which might have occurred at lower frequencies,
since it should have been detected by this complementary search.


Table~\ref{tab:obs} shows the amount of time in S2/DT8 during which 
each detector was operating.  As we shall see in Section~\ref{sec:tuning}, 
the LIGO-TAMA network achieved its lowest background rate during 
periods when both of the LIGO Hanford interferometers (H1 and H2) and
at least one of the LIGO Livingston and TAMA interferometers (L1 and
T1) were operating.
Restricting our analysis to these detector combinations gives us  
three independent data sets: the
quadruple-coincidence data set, denoted H1-H2-L1-T1; the data set
during which L1 was not operating, denoted H1-H2-nL1-T1, and the data
set during which T1 was not operating, denoted H1-H2-L1-nT1 (``n'' for
``not operating'').  The observation time for each of these data sets
is also shown in Table~\ref{tab:obs}.

The LIGO-TAMA quadruple-coincidence data set (H1-H2-L1-T1) is
particularly well-suited to making confident detections of
gravitational-wave bursts, since combining so many detectors
naturally suppresses the background from accidental coincidences 
-- to well below one per year, in our case -- while maintaining
high detection sensitivity.
Meanwhile, the triple-coincidence data sets (H1-H2-nL1-T1 and
H1-H2-L1-nT1) contribute the bulk of our observation time.  
In particular, the high T1 duty cycle (82\%)
allows us to use the large amount of H1-H2 data in H1-H2-nL1-T1
coincidence that would otherwise be discarded because of the
poor L1 duty cycle (33\%).
The LIGO-TAMA detector network therefore has more than twice
as much useful data as the LIGO detectors alone.
This increase in observation time allows a proportional
decrease in the limit on the GWB rate which we are able
to set with the combined detector network (for negligible background), and increases the 
probability of seeing a rare strong gravitational-wave event.
Furthermore, while the LIGO-TAMA network uses only half of 
the TAMA data, we shall see that the suppression of the 
background by coincidence allows it to place stronger upper limits
on weak GWBs than can TAMA alone.

\begin{table}[tbp]
  \begin{center}
    \begin{tabular}{|c r r|}
        \hline
        detector       &  observation &  fraction of total  \\ 
        combination    &  time (hr)   &  observation time \\ 
        \hline
        \hline
        H1             &  1040        &  74\%  \\ 
        H2             &   821        &  58\%  \\
        L1             &   536        &  38\%  \\ 
        T1             &  1158        &  82\%  \\  
        \hline
        \hline
        H1-H2-L1-T1    &   256        &  18\%  \\  
        H1-H2-nL1-T1   &   320        &  23\%  \\
        H1-H2-L1-nT1   &    62        &   4\%  \\
        \hline
        network totals &   638        &  45\%  \\
        \hline 
    \end{tabular}
    \caption{\label{tab:obs}
      Observation times and duty cycles of the LIGO and TAMA detectors 
      individually, and in various combinations, during S2/DT8.
      The symbol nL1 (nT1) indicates times when L1 (T1) was not operating.  
      The network data sets are disjoint (non-overlapping).
    }
  \end{center}
\end{table}

The LIGO and TAMA detectors had not yet reached their design
sensitivities by the time of the S2/DT8 run; nevertheless, the quantity
of coincident data available -- nearly 600 hours -- provided an excellent
opportunity to develop and test joint searches between our collaborations.
In addition, the sensitivity of these instruments in their common
frequency band was competitive with resonant-mass detectors (see for 
example \cite{Fa:04}), but with a much broader bandwidth.  Finally, 
there is always the possibility of a
fortunate astrophysical event giving rise to a detectable signal.


\section{Analysis Method}
\label{sec:method}

\begin{figure}[tbp] 
  \begin{center}
    \includegraphics[width=8cm]{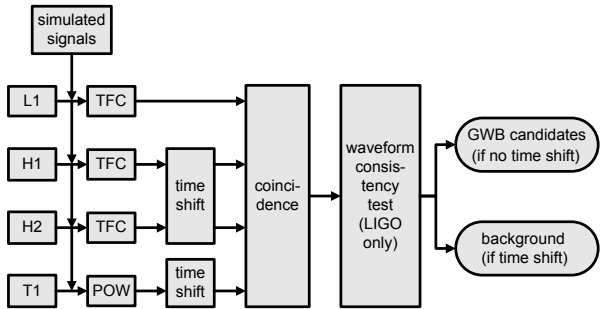}
    \caption{\label{fig:pipeline} 
      Schematic of our analysis pipeline.
      Data from each detector is analyzed for bursts using
      the TFClusters (TFC) or Excess Power (POW) algorithm.
      Optionally, a time shift of 5 - 115 s is added to the event triggers
      from some sites.  We look for simultaneous events from each
      operating detector, then apply the $r$-statistic waveform
      consistency test to the data from the LIGO detectors.
      Surviving coincidences are possible GWBs if no time shifts were
      used; otherwise they are accidental coincidences (background).  
      The detection efficiency of the network is estimated 
      by adding simulated GWBs to the data from each detector 
      and repeating the analysis.  Note that one of the
      L1 or T1 detectors may not be operating at any given time.    
    }
  \end{center}
\end{figure}

Our analysis methodology is similar, though not identical, to
that used in the LIGO S1 and S2 un-triggered GWB searches 
\cite{Ab_etal:05, Ab_etal:04}.  The essential steps are
illustrated in Figure~\ref{fig:pipeline}.  These are:
\begin{enumerate}
\item
Search the data from each detector separately for burst events.
\item 
Look for simultaneous (``coincident'') events in all operating detectors.
\item
Perform a waveform consistency test on the data from the LIGO
interferometers around the time of each coincidence.
\item 
Estimate the background rate from coincident detector noise fluctuations by
repeating the coincidence and waveform consistency tests after 
artificially shifting in time the events from different sites.
\item 
Compare the number of coincidences without time shifts 
to that expected from the background to set an upper limit 
on the rate of detectable bursts.
(A significant excess of events indicates a possible detection.)  
\item 
Estimate the network sensitivity to true 
GWBs (i.e., the false dismissal probability) 
by adding simulated signals to the detector
data and repeating the analysis.
\end{enumerate}  
In the following subsections we describe these steps in more detail.
In addition, the various thresholds used for event 
generation, coincidence, etc., are tuned to maximize the
sensitivity of the  analysis; this tuning is described in
Section~\ref{sec:tuning}.

The LIGO software used in this analysis is available publicly 
\cite{cite:CVS}.

\subsection{Event trigger generation}
\label{sec:etgs}

To maintain sensitivity to the widest range of signals, our 
burst-detection algorithms do not use templates.   
Instead, they look for transient excesses of power in the
detector output.  The production of lists of transient events,
or {\em event triggers}, was
done independently by LIGO and TAMA, using different algorithms.
Since both of the algorithms used have been described elsewhere, 
we review them only briefly.

\subsubsection{TFClusters}
\label{sec:tfcl}

The LIGO triggers were produced using the TFClusters burst-detection
algorithm \cite{Ab_etal:04,Sy:02}.

Before processing in TFClusters, the data from a given detector are
first high-pass filtered and whitened using a linear
predictor error filter \cite{ChBlMaKa:04,cite:LPEF_params}.  
The TFClusters algorithm then constructs a
time-frequency spectrogram of the filtered data by segmenting the
data into 50\% overlapping time intervals and Fourier transforming. 
The fraction $p$ of highest-power pixels in each frequency bin are 
selected as
{\em black} pixels, where $p$ is the {\em black pixel probability}.
(Note that this thresholding is inherently adaptive, so that the
rate of triggers is not unduly affected by slow trends in the noise
floor.)  Event triggers are formed from clusters of nearest-neighbor
black pixels that exceed a specified size. In keeping with our choice of
frequency band, only triggers that overlap 700-2000 Hz are retained; all
others are discarded.  These triggers are then passed to a function
which makes refined estimates of their peak time, duration, central
frequency, bandwidth, and signal-to-noise ratio.

\subsubsection{Excess Power}
\label{sec:power}

The TAMA triggers are generated using an excess power algorithm, 
following the procedure used in a TAMA-only search for GWBs \cite{An_etal:04}.  

The TAMA data are first conditioned to remove lines (including the
peaks at multiples of 400/3 Hz visible in Fig.~\ref{fig:spectra_zoom}).
It is then divided into 87.5\% overlapping segments and Fourier
transformed.  The resulting spectrogram is normalized by the background
estimated over the previous 30 min.  
The signal-to-noise ratio (SNR) is then summed over a fixed set of
frequency bins in the range 230-2500 Hz, and a trigger produced when
the SNR exceeds the threshold $\rho_0=4$.  Triggers separated by 
less than 25 ms are reported as a single event characterized by the 
peak time, duration, and SNR.  (Due to the use of a 
frequency mask, no frequency
information is assigned to the trigger.)  Finally, 
triggers occurring simultaneously with excursions in the intensity 
of the light in the recycling cavity are vetoed (ignored), as are 
triggers that fail a time-scale test designed to pass only 
millisecond-duration events \cite{An_etal:04}.

\subsection{Coincidence and background}
\label{sec:coinc_bckgrd}

To minimize the possibility of falsely claiming 
a gravitational-wave detection, we require any 
candidate GWB to be observed 
simultaneously by all operating detectors.
In this section we explain how the coincidence test was imposed, 
and how the background rate was estimated.

\subsubsection{Coincidence}
\label{sec:coinc}

The coincidence test is very simple.  Each event $i$ is characterized 
by a peak time $t_i$ and a duration $\Delta t_i$.  Events from two 
detectors are defined to be in coincidence if the difference in their 
peak times satisfies  
\be\label{eqn:coinc} 
\left|t_i-t_j \right| < w+\frac{1}{2}(\Delta t_i+\Delta t_j) \, . 
\ee 
Here $w$ is a coincidence ``window'' which accounts for 
the light travel time between the detectors in question; 
in practice we use windows 10 - 20 ms longer than the light travel time for
safety. 
The duration-dependent term allows for the estimated peak time of
coincident triggers to be farther apart if the triggers are long
compared to the coincidence window \cite{foot:trigdur}; 
one may consider this as an allowance for the uncertainty in the
determination of the peak time.  A set of event triggers $i, j,\ldots,
k$ is defined to be in coincidence if each pair $(i,j)$, $(i,k)$,
$(j,k)$, etc., is in coincidence.

Ideally, the window $w$ for each pair of detectors should be as short
as possible, to minimize the rate of accidental coincidences between
noise events in the various detectors, while still being long enough
that all simulated signals detected are in coincidence.   The
windows for our analysis are determined using the simulations
described in Section~\ref{sec:tuning}.

It is observed that triggers in the S2 and DT8 data tend to be
produced in clusters, on time scales of order 1 s or less.
We therefore count groups of coincident triggers
that are separated in time by less than 200 ms as 
a single GWB candidate when estimating the GWB rate.

Note that no attempt is made to compare the amplitude or SNR of
events between detectors.  Such comparisons are difficult due to the
differences in alignment of the detectors (except for the H1-H2
pair); see Figure~\ref{fig:antenna}.  
We do, however, impose a test on the consistency of
the waveform shape as measured by the various LIGO detectors; see
Section \ref{sec:rstat}.

\subsubsection{Background}
\label{sec:bckgrd}

\begin{figure}[tbp] 
  \begin{center}
    \includegraphics[width=8.5cm]{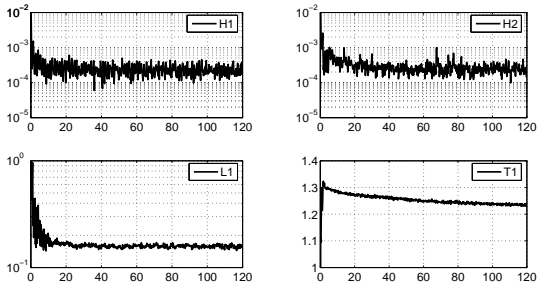}
    \caption{\label{fig:autocorr}
	Autocorrelogram of trigger peak times from each detector.
    These are histograms of the difference in peak time between 
    each pair of triggers from a given detector, binned in 0.2 s      
	intervals.  
    The horizontal axis is the time difference (units of s); the 
    vertical axis is the number of trigger pairs per bin divided 
    by the observation time and the bin width (units of s$^{-2}$).
    With this normalization, the mean value is the square of the 
    trigger rate.  
	An autocorrelogram value significantly above the mean indicates a 
	correlation in the time of event triggers.
    Each detector shows some excess above Poisson rates at delays 
    of up to a few seconds.  The sharp dip in the T1 curve near zero 
    time is due to the clustering of the T1 triggers. 
    The smaller relative fluctuations in the L1 and T1 curves are 
    due to the much higher trigger rates from these detectors, which 
    results in more triggers per bin.
    }
  \end{center}
\end{figure}

Even in the absence of real gravitational-wave signals, one expects
some coincidences between random noise-generated events.  We estimate
this background rate by repeating the coincidence procedure after
adding artificial relative time shifts of $\pm 5, 10, \ldots, 115$ s 
to the triggers from the LIGO Hanford and/or TAMA sites, 
as indicated in Figure~\ref{fig:pipeline}.  
(We do not shift the triggers from H1 and H2 relative to each
other, in case there are true correlated noise coincidences 
caused by local environmental effects.)
These shifts are
much longer than the light travel time between the sites, so that any
resulting coincidence cannot be from an actual gravitational wave.
They are also longer than detector noise auto-correlation times 
(see Figure~\ref{fig:autocorr}), and shorter than time scales 
on which trigger rates vary, so that each provides an independent 
estimate of the accidental coincidence rate.  

The H1-H2-nL1-T1 and H1-H2-L1-nT1 data sets each come from 2 sites, so
that we have 46 nonzero relative time shifts in  $\{-115, -110,
\ldots, 115\}$ s.  Hence, the smallest nonzero background rate that can be
measured for these data sets is approximately $(46T)^{-1}$,
where $T$ is the observation time \cite{foot:obstime}. 
The H1-H2-L1-T1 network has 3 sites, for a total of $47^2-1=2208$ 
independent time shifts.  We use all of these time shifts, so the
smallest nonzero background rate that we can measure for 
the quadruple-coincidence data set is approximately $(2208T)^{-1}$.

\subsection{Waveform consistency test}
\label{sec:rstat}

The event generation and coincidence procedures outlined above
are designed to detect simultaneous excesses of power in each
detector, without regard to the waveform of the event.
To test if the waveforms as measured in each detector are
consistent with one another (as one would expect for a GWB),
we apply a test based on the linear correlation coefficient
between data streams, the  $r$-statistic \cite{Ca:04}.
We will see in Section~\ref{sec:tuning} that the $r$-statistic test 
is very effective at eliminating accidental coincidences, with 
very little probability of rejecting a true
gravitational-wave signal.  
(See also \cite{Ab_etal:05} for demonstrations of the $r$-statistic 
with other simulated GWB waveforms.)

The $r$-statistic test consists of computing the cross-correlation
of the time-series data from pairs of detectors around the time
of a coincidence.  A GWB will increase the magnitude of the
cross-correlation above that expected from noise alone.
The measured cross-correlations are compared to those expected
from Gaussian noise using a Kolmogorov-Smirnov test with 95\%
confidence level.  If not consistent, then the logarithmic 
significance (negative log of the probability) of each 
cross-correlation is computed and averaged over detector pairs. 
We refer to the resulting quantity as $\Gamma$.
If the maximum averaged significance exceeds a threshold $\Gamma_0$,
then the coincidence is accepted as a candidate GWB; otherwise
it is discarded.
The threshold $\Gamma_0$ is chosen sufficiently high to reduce the 
background by the desired amount without rejecting too many
real GWB signals.  For more details on the test, see \cite{Ca:04}.

The $r$-statistic test was developed for use in LIGO searches, and
it is based on the premise that a real gravitational-wave signal will
have similar form in different detectors.
It is not clear that it can be applied safely to
detectors with very different orientations (such as LIGO and TAMA),
which see different combinations
of the two polarizations of a gravitational wave.
Since this matter is still under study, we
use the $r$-statistic test to compare data between the
LIGO detectors only (i.e., H1-H2, H2-L1, and L1-H1, 
but not including T1).

\subsection{Statistical analysis}
\label{sec:stats}

Our scientific goal of this search is to detect GWBs, or in the
absence of detectable signals, to set an upper limit on their
mean rate, and to estimate the minimum signal amplitude to which
our network is sensitive.

The coincidence procedure described in Section~\ref{sec:coinc_bckgrd}
produces two sets of coincident events.  The set with no artificial
time shift is produced by background noise and possibly also by
gravi\-ta\-tional-wave bursts.  The time-shifted set contains only
events produced by noise, and hence characterizes the background.

Given the number of candidate GWBs and the estimate of the number of
accidental coincidences expected from the background, we use the
Feldman-Cousins technique \cite{FeCo:98} to compute the 90\%
confidence level upper limit or confidence interval on the rate of
detectable gravi\-ta\-tional-wave bursts.  In practice, since we are not
prepared to claim a detection based only on such a statistical
analysis, we choose in advance to use only the upper value of the
Feldman-Cousins confidence interval.  We report this upper value $R_{90\%}$
as an upper limit on the GWB rate, regardless of whether
the Feldman-Cousins confidence interval is consistent with a rate of zero.  
Because of this modification our upper limit procedure has 
a confidence level greater than 90\%; i.e., our upper limits 
are conservative.

The rate upper limit $R_{90\%}$ from the Feldman-Cousins procedure
applies to GWBs for which our network has perfect detection
efficiency.  For a population of GWB sources for which our detection
efficiency is $\epsilon(h)$, where $h$ is the GWB amplitude and
$0\le\epsilon(h)\le1$, the corresponding rate upper limit
$R_{90\%}(h)$ is
\be\label{eqn:R}
R_{90\%}(h) \le \frac{R_{90\%}}{\epsilon(h)} \, .
\ee 
This defines a region of rate-versus-strength space which 
is excluded at 90\% confidence by our analysis.  The exact domain 
depends on the signal type through our efficiency $\epsilon(h)$.  
We will construct such exclusion regions for one hypothetical 
population of GWB sources.


\section{Simulations and Tuning}
\label{sec:sims_tuning}

There are a number of parameters in the analysis pipeline of 
Figure~\ref{fig:pipeline} that can be manipulated to adjust 
the sensitivity and background rate of our network.
The most important are the thresholds for trigger generation (the 
TFClusters black pixel probability $p$ and the Excess Power SNR threshold 
$\rho_0$), the $r$-statistic threshold $\Gamma_0$, and the coincidence 
windows $w$ for each detector pair.  
Our strategy is to tune these parameters to maximize the sensitivity 
of the network to millisecond-duration signals while maintaining 
a background of less than 0.1 surviving coincidences
expected over the entire S2/DT8 data set.

\subsection{Simulations}
\label{sec:sims}

The LIGO-TAMA network consists of widely separated
detectors with dissimilar noise spectra and antenna responses.
To estimate the sensitivity of this heterogeneous network we add (or
``inject'') simulated gravi\-ta\-tional-wave signals into the data streams
from each detector, and re-analyze the data in exactly the same manner
as is done in the actual gravitational-wave search; this is indicated
in Figure~\ref{fig:pipeline} by the ``simulated signals'' box.    
These injections are done coherently; i.e., they correspond to a
GWB incident from a specific direction on the sky.  The simulated
signals include the effects of the antenna response of the detectors,
and the appropriate time delays due to the physical separation of the detectors. 

These simulations require that we specify a target population,
including the waveform and the distribution of sources over the sky.
We select a family of simple waveforms that have millisecond durations 
and that span the frequency range of interest, 700-2000 Hz.
Specifically, we use linearly polarized Gaussian-modulated sinusoids:
\bea\label{eqn:h}
h_+(t) 
  & = &  h_{\ind{rss}} \left( \frac{\pi}{2f_0^2}\right)^{-1/4}
         \!\!\!\!
         \sin{\left[2\pi f_0(t-t_0)\right]} \, 
         {\rm e}^{-\frac{(t-t_0)^2}{\tau^2}} \, , \hphantom{0.2in} \\
h_\times(t) 
  & = &  0 \,. \nonumber
\eea
(Other waveforms, along with these, have been considered  
in \cite{Ab_etal:05,An_etal:04,Ab_etal:04}.)  
Here $t_0$ is the peak time of the signal envelope.  
The central frequency $f_0$ of each injection is picked randomly from
the values 700, 849, 1053, 1304, 1615, 2000 Hz, which span our analysis
band in logarithmic steps.  The efficiency of detection of these
signals  thus gives us a measure of the network sensitivity averaged
over our band.   We set the envelope width as $\tau = 2/f_0$,  which
gives durations of approximately 1-3ms.   The corresponding quality
factor is $Q\equiv\sqrt{2}\pi f_0\tau=8.9$ and the bandwidth is
$\Delta f = f_0/Q \simeq 0.1f_0$, so these are narrow-band  signals.

The quantity $h_{\ind{rss}}$ in equation~(\ref{eqn:h}) is the
root-sum-square amplitude of the plus polarization:
\be\label{eqn:hrss}
\left[ \int_{-\infty}^\infty \!\!\! dt \, h^2_+(t) \right]^{1/2}
  =  h_{\ind{rss}} 
\ee
We find $h_{\ind{rss}}$ to be a convenient measure of the signal strength.
While it is a detector-independent amplitude,
$h_{\ind{rss}}$ has the same units as the strain noise amplitude 
spectrum of the detectors, which allows for a direct comparison of the 
signal amplitude relative to the detector noise.
All amplitudes quoted in this report are $h_{\ind{rss}}$ amplitudes.

Lacking any strong theoretical bias for probable sky positions of
sources of short-duration bursts, we distribute the simulated signals
isotropically over the sky. 
We select the polarization angle randomly with uniform distribution 
over $[0,\pi]$.

A total of approximately 16800 of these signals were injected 
into the S2/DT8 data.  
For each signal, the actual waveform $h(t)$ as it would be seen
by a given detector was computed,
\be\label{eqn:h_of_t}
h(t) = F^+ h_+(t) + F^\times h_\times(t) =  F^+ h_+(t) \, ,
\ee
and $h(t)$ was added to the detector data.  Here $F^+$, $F^\times$
are the usual antenna response factors, which are functions of the
sky direction and polarization of the signal relative to the detector
(see for example \cite{AnBrCrFl:01}).  The signals in the different
detectors were also delayed relative to one another  according to the
sky position of the source.

These simulated signals were shared between LIGO and TAMA by writing
the signals $h(t)$ in frame files \cite{frames,MDC}, including the
appropriate detector response and calibration effects.  
These signal data were added to the data streams from the individual
detectors before passing through TFClusters or Excess Power.
In addition to providing estimates of the network detection
efficiency, the ability of these two independent search codes
to recover the  injected signals is an important test of the
validity of the pipeline.

An injected signal is considered detected if there is a coincident
event from the network within 200 ms of the injection time.  The
network efficiency $\epsilon(h_{\ind{rss}})$ is simply the fraction of events of
amplitude $h_{\ind{rss}}$ which are detected by the network.  We find 
that good empirical fits to the measured efficiencies can be found 
in the form 
\begin{equation}\label{eq:fit}
\epsilon(h_{\ind{rss}})
  =  \frac{1}{
         \displaystyle
         1+\left(\frac{h_{\ind{rss}}}{h_{\ind{rss}}^{50\%}}\right)^{
             \alpha\left[1+\beta\tanh\left(h_{\mathrm{rss}}/h_{\ind{rss}}^{50\%}\right)\right]
         }
     } \, ,
\end{equation} 
where $h_{\ind{rss}}^{50\%} > 0$, $\alpha < 0$, and $-1 < \beta \le 0$.  
Here $h_{\ind{rss}}^{50\%}$ is the amplitude at which
the efficiency is 0.5,  
$\alpha$ parameterizes the width of the transition 
region, and $\beta$ parameterizes the asymmetry of the 
efficiency curve about $h_{\ind{rss}}=h_{\ind{rss}}^{50\%}$. 
When presenting efficiencies we will use fits of this type.

As we shall see, the efficiency
transitions from zero (for weak signals) to unity (for
strong signals), over about an order of magnitude in signal
amplitude.  It proves convenient to characterize the network
sensitivity by the single number $h_{\ind{rss}}^{50\%}$ 
at which the efficiency is 0.5.  This amplitude is a
function of the trigger-generation thresholds; it and the background rate are
the two performance measures that we use to guide the tuning of our 
analysis.

\subsection{Tuning procedure}
\label{sec:tuning}

\begin{figure}[tbp] 
  \begin{center}
  \includegraphics[width=8.5cm]{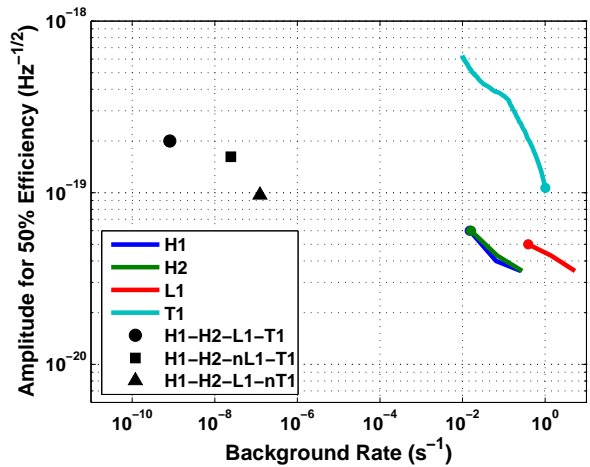}
  \caption{
  \label{fig:oper}
  Amplitude $h_{\ind{rss}}^{50\%}$ for 50\% detection efficiency versus
  background rate for each of the individual detectors, and for the
  three coincidence combinations.  The circles on the single-detector 
  curves indicate the tuning selected for trigger generation.
  The circle, square, and triangle denote the resulting
  amplitude at 50\% efficiency and
  an upper limit on the background rate for the H1-H2-L1-T1, H1-H2-nL1-T1, and
  H1-H2-L1-nT1 networks after the $r$-statistic with these tuning
  choices.  (We can only compute upper limits on the background rates for 
  coincidence because no time-shifted coincidences survive the waveform
  consistency test.)  The efficiency is averaged over all of the
  sine-Gaussian signals in our analysis band.}
\end{center}
\end{figure}

As stated earlier, our tuning strategy is to maximize the detection
efficiency of the network while maintaining a background rate of less than
approximately 0.1 events over the entire data set.
For simplicity, we
chose a single tuning for the production and analysis of all event
triggers from all data sets.  This strategy is implemented as follows:
\begin{enumerate}
\item
For TFClusters, the efficiency for detecting the sine-Gaussian 
signals and the background rate are measured for each detector 
for a large number of parameter choices.  
For each black-pixel probability $p$ (which determines the background rate) 
the other ETG parameters are set to obtain the lowest $h_{\ind{rss}}^{50\%}$ 
value \cite{foot:TFCparams}. The TAMA
Excess Power algorithm is tuned independently for short-duration
signals as described in \cite{An_etal:04}.
The resulting 
performance of each detector is shown in Figure~\ref{fig:oper}.
\item
The coincidence window $w$ for each detector pair 
in equation (\ref{eqn:coinc}) is fixed by
performing coincidence on the triggers from the simulated signals.  We
find that selecting windows only slightly larger (by $\sim1$ms)  than
the light travel time  between the various detector pairs (see
Table~\ref{tab:detsep}) ensures that all of  the
injections detected by all interferometers produce coincident
triggers.
For simplicity, we use a single window of $w=20$ ms for coincidence
between any LIGO detectors \cite{foot:H1H2window}
and a single window of $w=43$ ms for coincidence between any LIGO detector 
and TAMA.  These choices correspond to using the longest possible 
time delay plus a $\sim$10 ms safety margin.
\begin{table}[tbp]
  \begin{center}
    \begin{tabular}{|l r r|}
      \hline
      detector pair  & separation (km)  & separation (ms)  \\
      \hline
      LHO-LLO  &  3002  &  10.0  \\
      LLO-TAMA &  9683  &  32.3  \\
      TAMA-LHO &  7473  &  24.9  \\
      \hline
    \end{tabular}
    \caption{\label{tab:detsep}
      Separation of the LIGO and TAMA interferometers, using data from 
      \cite{AnBrCrFl:01}.  
    }
  \end{center}
\end{table}
\item
To obtain the best network sensitivity versus background rate, we select
the single-detector ETG thresholds ($p, \rho_0$) to match
$h^{50\%}_{\ind{rss}}$  as closely as possible between the detectors.  (This
is similar in spirit to the IGEC tuning \cite{As_etal:03}, although not
the same, as we are not able to easily compare the amplitude of
individual events from our misaligned broadband detectors.)
In practice, the TAMA detector has slightly poorer sensitivity than
the LIGO detectors.  We therefore set the TAMA threshold as low as
we consider feasible, $\rho_0=4$;  this sets the sensitivity of the
network as a whole.  We then choose the LIGO single-detector
thresholds for similar efficiency.  
\item
The final threshold is that for the $r$-statistic, denoted $\beta$.   In
practice we find that the $r$-statistic has negligible effect on the
network efficiency for $\beta<5$.  We set $\beta=3$, which proved 
sufficient to eliminate all time-lagged (accidental) coincidences 
while rejecting less than 1\% of the injected signals.
\end{enumerate}
Figure~\ref{fig:oper} shows the 
resulting $h_{\ind{rss}}^{50\%}$ and background rate for each of the 
three coincidence combinations.  Table~\ref{tab:results} shows trigger 
rates and livetimes for each coincidence combination.

We note from Figure~\ref{fig:oper} that the network background rates are 5-9 
orders of magnitude smaller than the rates of the individual detectors.  
Roughly speaking, adding a detector with event rate $R_i$ and coincidence 
window $w$ to the network changes the network background rate by a factor of
approximately $2R_iw$.
From the single-detector rates of Figure~\ref{fig:oper} 
or Table~\ref{tab:results} 
we estimate that 
H1 and H2 each reduce the background rate by $\sim10^{3}$, L1 by 
$\sim50$, and T1 by $\sim10$. 
This is why we require both H1 and H2 to be 
operating: they suppress strongly the background for our network.  

We have confirmed that the background rate estimated from time
shifts is consistent with that expected from Poisson statistics.  
Assuming Poisson statistics, the expected background rate $R$ for a set of 
$N$ detectors with rates $R_i$ is approximately 
\begin{equation}
\label{eqn:poissonrate}
R \approx \frac{1}{w} \prod_{i=1}^N 2 R_i w
\end{equation}
where we assume a single coincidence window $w$ for simplicity.  
Using this formula and the single-detector rates from 
Table~\ref{tab:results}, one predicts background 
rates before the r-statistic consistent with those 
determined from time delays.  This agreement gives increased 
confidence in our background estimation.

It is also worth noting that the 50\% efficiency point $h_{\ind{rss}}^{50\%}$ 
is a very shallow function of the background rate for multiple
detectors.  Hence, there is little value in lowering the trigger 
thresholds to attempt to detect weaker signals.  For example,
allowing the triple-coincidence background rate of TFClusters (the rate for 
the H1-H2-L1-nT1 data) to increase by 3 orders of magnitude lowers 
$h_{\ind{rss}}^{50\%}$ by less than a factor of 2.  
For four detectors, $h_{\ind{rss}}^{50\%}$ varies even more slowly 
with the background rate.  This is why we tune for $\ll$1 background 
event over the observation time;  there is almost no loss of efficiency 
in doing so.

To avoid bias from tuning our pipeline using the same
data from which we derive our upper limits, the tuning was done
without examining the full zero-time-shift coincidence trigger sets.
Instead, preliminary tuning was done using a 10\% subset of the
data, referred to as the {\em playground}, which was not used for
setting upper limits.  Final tuning choices were made by examining
the time-shifted coincidences and the simulations over the full data
set.  As it happens, the only parameter adjusted in this final tuning 
was the $r$-statistic threshold $\Gamma_0$; we required the full
observation time to have enough background coincidences  
to allow reasonably accurate estimates of the background suppression 
by the $r$-statistic test.  

\begin{figure}[tbp] 
  \begin{center} 
  \includegraphics[width=8.5cm]{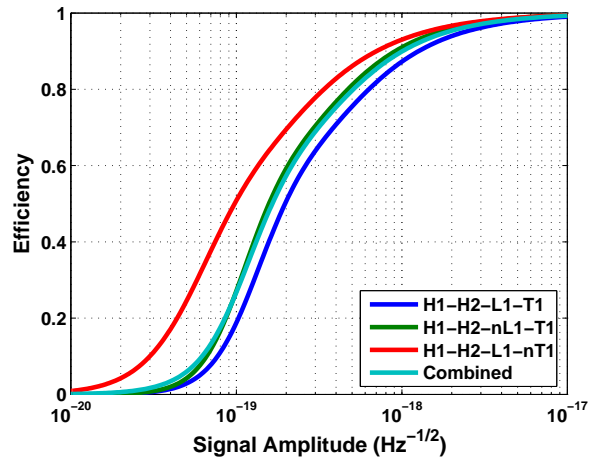}
  \caption{
  \label{fig:sigmoid}
  Detection efficiency over the various data sets individually,  and
  combined, using the final tuning.  The combined efficiency curve is
  the average of the curves  for the three data sets, weighted by their
  observation times.  These efficiencies are averaged over all of the
  sine-Gaussian signals in our analysis band.  There is a statistical
  uncertainty at each point in these curves of approximately 1-3\% 
  due to the finite number of simulations performed.  
  }
  \end{center}
\end{figure}

Figure~\ref{fig:sigmoid} shows the efficiency of the LIGO-TAMA 
network as a function of signal amplitude for each of the three
data sets, and also the average efficiency weighted by the
observation time of each data set.  By design, the efficiencies
are very similar, with $h_{\ind{rss}}^{50\%}$
values in the range $1$-$2\times10^{-19}$ Hz$^{-1/2}$.  

\begin{figure}[tbp] 
  \begin{center}
  \includegraphics[width=8.5cm]{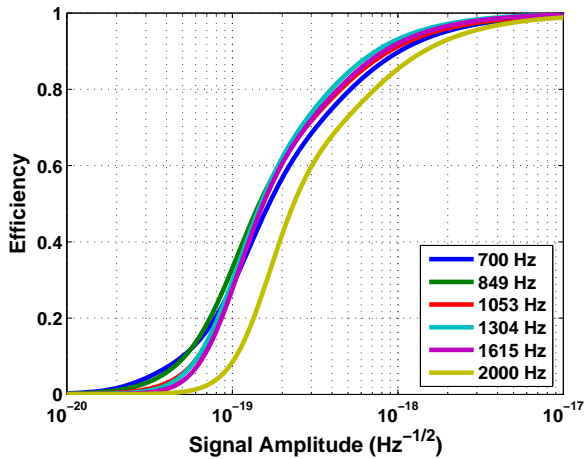}
  \caption{
  \label{fig:sigmoid2}
  Detection efficiency for the combined data set, by central 
  frequency $f_0$ of the sine-Gaussian signal in equation (\ref{eqn:h}). 
  There is a statistical uncertainty at each point in these curves 
  of approximately 2-4\% due to the finite number of simulations performed.
  }
  \end{center}
\end{figure}

\begin{figure}[tbp] 
  \begin{center}
  \includegraphics[width=8.5cm]{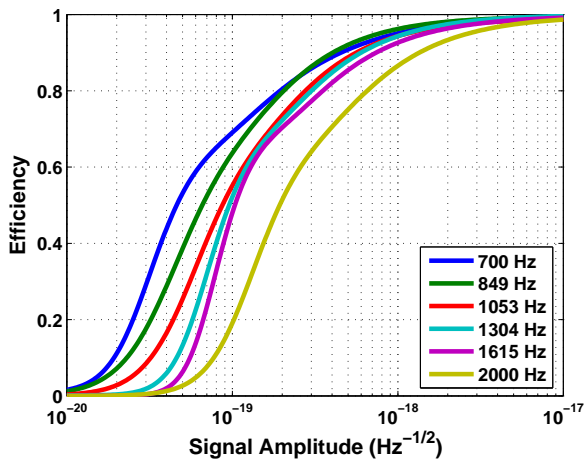}
  \caption{
  \label{fig:sigmoid3}
  Detection efficiency for the H1-H2-L1-nT1 data set (i.e., with only the 
  LIGO detectors operating), by central frequency $f_0$ of the sine-Gaussian 
  signal in equation (\ref{eqn:h}).
  There is a statistical uncertainty at each point in these curves 
  of approximately 2-6\% due to the finite number of simulations performed.
  The improved efficiency for lower-frequency signals indicates 
  that sensitivity at these frequencies is limited by the TAMA 
  detector.  This behavior is consistent with the noise spectra
  shown in Figure~\ref{fig:spectra_zoom}.
  }
  \end{center}
\end{figure}

Figure~\ref{fig:sigmoid2} shows how the combined efficiency varies
across our frequency band; the weak dependence on the central 
frequency of the injected signal is a
consequence of the flatness of the envelope of the detector noise
spectra shown in Figure~\ref{fig:spectra_zoom}.  This is corroborated
by the efficiency for the H1-H2-L1-nT1 data set (without TAMA), shown
in Figure~\ref{fig:sigmoid3}.  The improvement in the low-frequency
sensitivity for this  data set indicates that TAMA limits the network
sensitivity at low frequencies, as expected from the noise spectra.


\subsection{Systematic and statistical uncertainties}
\label{sec:uncert}

The only significant systematic uncertainty in our analysis 
is in the overall multiplicative scale of the calibration (the  
coupling of strain to the output of the individual detectors).  
The ``1-$\sigma$'' uncertainties were estimated as $\sim$9\%
for L1 and $\sim$4\% for each of H1, H2, and T1 \cite{S2cal}.
Simple Monte-Carlo modeling indicates that, with 90\% confidence, 
the $h_{\ind{rss}}^{50\%}$ 
value for any given network combination will not be more than 4\% larger 
than the estimated value due to these uncertainties.
We allow for this 
uncertainty in our rate-versus-strength plots by shifting our 
limit curves to larger $h_{\ind{rss}}$ by 4\%.

The main statistical uncertainty in our results is in the efficiency 
at any given signal amplitude, due to the finite number of simulations 
performed.  This can be quantified through the uncertainty in the 
parameters found for the efficiency fits (\ref{eq:fit}), 
and is typically less 
than 5\%.  We account for this by shifting our rate-versus-strength upper 
limit curves upward at each amplitude by 1.28 times the estimated statistical 
uncertainty in the corresponding efficiency.    
(The factor 1.28 gives a 90\% limit, assuming Gaussian statistics.)


\section{Analysis Results}
\label{sec:results}

After making the final tuning choices, we performed the  coincidence
analysis without time shifts for all three data sets.   No event
triggers survived the coincidence and $r$-statistic tests, so we
have no candidate gravitational-wave signals.

Table~\ref{tab:results} shows for each data set the rate  of triggers,
the number of coincident events before and after the $r$-statistic
test, and the total amount of data analyzed after  removing the
playground and accounting for the dead time of the TAMA  vetoes.  Also
shown are the number of accidental coincidences and  the effective
observation time from the time-shift experiments,  which provide our
estimate of the background rates.  Finally, the upper limits on the
rate of  detectable gravitational-wave bursts are shown.

As discussed in Section~\ref{sec:stats}, our upper limits are obtained 
using the Feldman-Cousins procedure \cite{FeCo:98}.  
This algorithm compares the observed number of events 
to that expected from the background.  As a rule, for 
a fixed number of observed events, the upper limit is 
stronger (lower) for higher backgrounds.  Since our backgrounds 
are too low to be measured accurately (there are 
no surviving time-shifted coincidences after the $r$-statistic), 
we conservatively assume zero background in calculating our upper limits.  
Since there are also no surviving coincidences without 
time shifts, the rate limits from the Feldman-Cousins 
procedure take on the simple form 
\be\label{eqn:ul}
R^i_{90\%} = \frac{2.44}{T_i}
\ee
where $T_i$ is the observation time for a particular network combination 
(see Table IV of \cite{FeCo:98} with $b=0, n=0$).
This gives the limits shown in Table~\ref{tab:results}.  
Additionally, since all three data sets have essentially 
zero background, we can treat them collectively 
as a single experiment by summing their observation times 
and the number of detected events (which happens to be zero):
\be\label{eqn:ul_comb}
R^{\ind{combined}}_{90\%} = \frac{2.44}{\sum_i T_i}
\ee
The resulting upper limit of 0.12 detectable events per day at 
90\% confidence is the primary scientific result of this analysis.

\begin{figure} 
  \begin{center}
  \includegraphics[width=8.5cm]{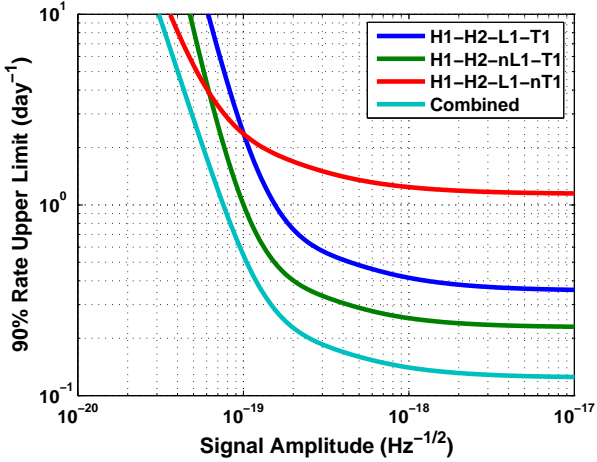}
  \caption{
  \label{fig:ul}
  Rate-versus-strength upper limits from each LIGO-TAMA data set, and
  combined, for the isotropic distribution of sources of sine-Gaussian 
  GWBs described in Section~\ref{sec:sims}.  The region above 
  any curve is excluded by that experiment with at least
  90\% confidence.  These curves include the allowances for  
  uncertainties in the calibration and in the efficiencies  
  discussed in Section~\ref{sec:uncert}.
  }
  \end{center}
\end{figure}

By dividing the rate upper limits by the efficiency 
for a given population of GWB sources, as in equation 
(\ref{eqn:R}), we obtain upper 
limits on the GWB rate as a function of the burst amplitude.  
Averaging over the network combinations gives
\be\label{eqn:Rvsh_comb}
R^{\ind{combined}}_{90\%}(h_{\ind{rss}})
  =  \frac{2.44}{\sum_i \epsilon_i(h_{\ind{rss}}) T_i}
\ee
For example, for our tuning population of isotropically 
distributed sources of sine-Gaussian GWBs, and averaging over 
all $f_0$ (i.e., using the efficiencies in Figure~\ref{fig:sigmoid}), 
one obtains the rate-versus-strength upper limits shown in Figure~\ref{fig:ul}.  
GWB rates and amplitudes above a given curve are excluded by
that data set with at least 90\% confidence.

\begin{table*}[tbp]
  \begin{center}
    \begin{tabular}{| l | c | c | c | c |}
      \hline 
      Data Set       &  H1-H2-L1-T1 &  H1-H2-nL1-T1 &  H1-H2-L1-nT1 & Combined \\
      \hline
      $R_{\ind{H1}}$ (s$^{-1}$)  &  0.0157              &  0.0151             &  0.0137               &         \\
      $R_{\ind{H2}}$ (s$^{-1}$)  &  0.0164              &  0.0183             &  0.0150               &         \\
      $R_{\ind{L1}}$ (s$^{-1}$)  &  0.399               &  -                  &  0.377                &         \\
      $R_{\ind{T1}}$ (s$^{-1}$)  &  1.03                &  1.04               &  -                    &         \\
      $N$                        &  0/0                 &  1/0                &  0/0                  & 1/0     \\
      $T$ (hr)                   &  165.3               &  257.0              &  51.2                 & 473.5   \\ 
      \hline
      $N_{\ind{bck}}$            &  31/0                &  57/0               &  0/0                  &  \\
      $T_{\ind{bck}}$ (hr)       &  3.422$\times10^5$   &  1.139$\times10^4$  &  2.243$\times10^3$    &  \\
      $\langle N \rangle$        &  0.015/$<$0.0005     &  1.3/$<$0.03        &  $<$0.03/$<$0.03      &  $<$1.4/$<$0.05     \\
      \hline
      $R_{90\%}$ (day$^{-1}$)    &   0.35               &  0.23               &  1.1                  &  0.12 \\
      \hline
    \end{tabular}
    \caption{\label{tab:results}  Results of the LIGO-TAMA analysis 
    for each data
    set separately, and combined.  $R_{\ind{H1}}$, etc., are the measured
    single-detector trigger rates.  $N$ is the total number of
    coincidences before/after the $r$-statistic waveform consistency test.  $T$ is the total
    observation time analyzed, after removal of the playground and
    veto dead times.  $N_{\ind{bck}}$ and $T_{\ind{bck}}$ are the corresponding
    summed  numbers from the  time-shift experiments.  $\langle N
    \rangle$ is the expected number of accidental  coincidences during
    the observation time.  (For $N_{\ind{bck}}=0$, we estimate $\langle N
    \rangle<T/T_{\ind{bck}}$.)  $R_{90\%}$ is the resulting upper limit on
    the rate of detectable  gravitational-wave events, at 90\%
    confidence.
    }
  \end{center}
\end{table*}

\subsection{Comparison to other searches}

The LIGO-TAMA search for GWBs is one of several such searches 
reported recently.  Table~\ref{tab:comparison} shows the observation 
time, rate upper limit, and approximate frequency band for 
LIGO-TAMA, the LIGO-only S2 search \cite{Ab_etal:05}, the TAMA-only 
DT9 search \cite{An_etal:04}, and the IGEC 
search \cite{As_etal:03}.
Our limit of 0.12 events per day is the strongest limit yet placed 
on gravitational-wave bursts by broadband detectors.  Even so, it is 
still approximately a factor of 30 larger than the IGEC limit, 
which was derived from approximately two years of data from a 
network of 5 resonant-mass detectors.  
Note however that the broadband nature of the LIGO and TAMA detectors 
means that they are sensitive to a wider class of signals than resonant-mass 
detectors; the IGEC search is only sensitive to GWBs with significant power at 
the resonant frequencies of all of the operating detectors.

\begin{table}[tbp]
  \begin{center}
    \begin{tabular}{| l | c | c | c |}
    \hline
    Network           &  T (day)  &  $R_{90\%}$ (day$^{-1}$) & band (Hz) \\
    \hline
    LIGO-TAMA S2/DT8  &   19.7    &  0.12       &  700-2000  \\
    LIGO-only S2      &   10.0    &  0.26       &  100-1100  \\
    TAMA-only DT9     &    8.1    &  0.49       &  230-2500  \\
    IGEC              &  707.9    &  0.0041     &  694-930   \\
    \hline
    \end{tabular}
    \caption{\label{tab:comparison}
    Observation times, rate upper limits, and frequency bands 
    for LIGO-TAMA and other recent burst searches 
    \cite{Ab_etal:05,As_etal:03,An_etal:04}.  
    The stated frequency range for the IGEC search is the range of
	the resonant frequencies of the detectors used.  The IGEC upper
	limit applies only to signals 
    with significant power at 
    the resonant frequencies of all of the operating detectors.  
    }
  \end{center}
\end{table}

\begin{figure} 
  \begin{center}
  \includegraphics[width=8.5cm]{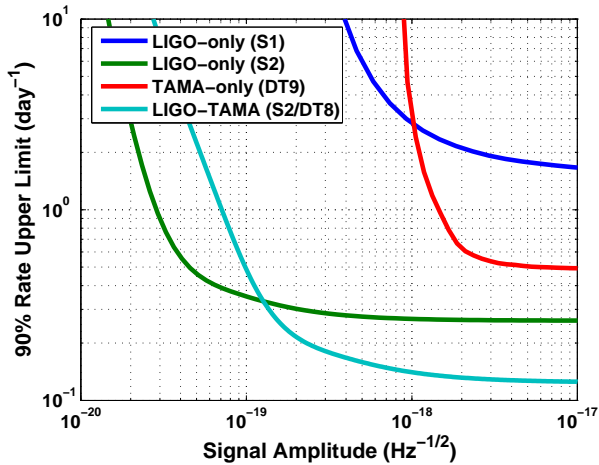}
  \caption{
  \label{fig:ul3}
  Comparison of the rate-versus-strength upper limits for $f_0=849$ Hz 
  sine-Gaussians from the combined LIGO-TAMA data set (including 
  systematic and statistical uncertainties) with those from 
  the LIGO-only S1 and S2 bursts searches \cite{Ab_etal:04,Ab_etal:05} 
  and the TAMA-only DT9 search \cite{An_etal:04}.
  The combined LIGO-TAMA network has a superior rate upper limit for strong 
  signals due to its larger observation time, while the LIGO-only S2 
  network has better sensitivity to weak signals.
  The TAMA-only DT9 amplitude sensitivity is limited by the 
  high SNR threshold needed achieve a background 
  rate of order one event over the observation time.
  Note that the LIGO-only S2 search had a nominal frequency range 
  of 100-1100 Hz, while the LIGO-TAMA search band is 700-2000 Hz.
  }
  \end{center}
\end{figure}

Since our sine-Gaussian test waveforms are narrow-band signals, we cannot 
compare directly our sensitivity to that of the IGEC network.
More concrete comparisons can be made between the performance of
the LIGO-TAMA network and LIGO and TAMA individually, by considering
the rate-versus-strength upper limit for $f_0=849$ Hz sine-Gaussians.  
Figure~\ref{fig:ul3} shows the upper limits for this waveform from 
the LIGO-only S1 and S2 searches \cite{Ab_etal:04,Ab_etal:05},
the TAMA DT9 search \cite{An_etal:04}, and the present analysis.
Compared to LIGO alone, the much longer observation
time afforded by joining the LIGO and TAMA detectors
allows the LIGO-TAMA network to set stronger rate upper
limits for amplitudes at which both LIGO and TAMA are sensitive.  
The joint network also enjoys a lower background rate from accidental
coincidences, particularly for the quadruple-coincidence network:
of order 1/40 yr$^{-1}$ for quadruple coincidence, versus of
order 2 yr$^{-1}$ for the LIGO-only S2 analysis.
However, the band of good sensitivity for LIGO-TAMA does not 
extend to low frequencies, due to the poorer TAMA noise level there.
The LIGO-only analysis also has better sensitivity
to weak signals, especially near the lower edge of our frequency band. 
Compared to TAMA alone, the LIGO-TAMA network has better sensitivity
to weak signals because coincidence with LIGO lowers the network
background rate without requiring high thresholds for trigger generation. 
For example, while the TAMA noise levels were lower in DT9 than in DT8,
the TAMA DT9 amplitude sensitivity is not as good as that of LIGO-TAMA 
due to the need to use a very high SNR threshold -- $\rho_0 = 10^4$ --
to reduce the TAMA-only background rate to of order one
event over the observation time.

\section{Conclusion}
\label{sec:summary}

The LIGO and TAMA collaborations have completed their first  
joint search for gravitational-wave bursts,
using 473hr of coincident data collected during early 2003.
We looked for millisecond-duration gravitational-wave 
bursts in the frequency range 700-2000 Hz, where all four of the 
detectors had comparable sensitivity.  
To maintain a low background, we analyzed data only 
from periods when at least three interferometers (including the 
two LIGO-Hanford interferometers) were 
operating, and we required candidate signals to be observed 
simultaneously in all of the operating detectors.  
We used coordinated injections of simulated gravitational-wave 
signals to estimate the detection efficiency of our heterogeneous
network.  We matched the efficiency between detectors to maximize
the network sensitivity while limiting the background rate to
less than 0.1 events expected over the entire observation time.
No gravitational-wave candidates were observed, and 
we place an upper bound of 0.12 events per day on the 
rate of detectable millisecond-duration gravitational-wave bursts  
with at least 90\% confidence. 
Simulations indicate that our network has a detection 
efficiency of at least 50\% (90\%) for narrow-band signals  
with root-sum-square strain amplitude greater than approximately 
$2\times10^{-19}$ Hz$^{-1/2}$ 
($10^{-18}$ Hz$^{-1/2}$) in the frequency band 700-2000 Hz.

This analysis highlights both advantages and disadvantages 
of joint coincidence searches compared to independent searches 
by LIGO and TAMA.  Together, the LIGO-TAMA network 
has more than twice as much data with three or more detectors in 
simultaneous operation than LIGO alone, leading to 
stronger rate limits.  We also enjoy a background rate of order
one event per 40 years (or lower) in quadruple-coincidence operation.
The lower background from coincidence also allows the TAMA data 
to be analyzed with lower thresholds for signal detection. 
These benefits come at some cost in detection efficiency and in 
bandwidth, particularly at low frequencies.
This is a result of requiring coincident detection by all 
interferometers, in which case the network sensitivity is
limited by the least sensitive detector at each frequency.

This analysis may serve as a prototype for more comprehensive 
collaborative searches in the future.  One improvement would 
be to expand the detector network.  For example, GEO, LIGO, 
and TAMA performed coincident data taking during Oct.~2003 - Jan.~2004;
a GEO-LIGO-TAMA network would contain 5 interferometers 
at four sites, with excellent sky coverage.
Another improvement would be to implement a fully coherent 
consistency test of coincident events, including all of the 
detectors in the network.  For example, the Gursel-Tinto 
technique \cite{GuTi:89} would allow us to take advantage of 
the different detector orientations to try to extract sky 
direction and waveform information from detected 
gravitational-wave signals.  It would also allow us to 
reject a coincidence if no consistent sky direction or 
waveform could be determined \cite{We:04}.

\section{Acknowledgments}

The LIGO scientific collaboration gratefully acknowledges 
the support of the United States
National Science Foundation for the construction and operation of
the LIGO Laboratory and the Particle Physics and Astronomy Research
Council of the United Kingdom, the Max-Planck-Society and the State
of Niedersachsen/Germany for support of the construction and
operation of the GEO600 detector. The authors also gratefully
acknowledge the support of the research by these agencies and by the
Australian Research Council, the Natural Sciences and Engineering
Research Council of Canada, the Council of Scientific and Industrial
Research of India, the Department of Science and Technology of
India, the Spanish Ministerio de Educaci\'on y Ciencia, the John Simon
Guggenheim Foundation, the Leverhulme Trust, the David and Lucile
Packard Foundation, the Research Corporation, and the Alfred P.
Sloan Foundation.
TAMA research is supported by a Grant-in-Aid for Scientific Research on 
Priority Areas (415) of the Japanese Ministry of Education, Culture, 
Sports, Science, and Technology.
This document has been assigned LIGO Laboratory document number
LIGO-P040050-06-Z.



\end{document}

%% file: ligotama_auth_v5.tex



\newcommand*{\AG}{Albert-Einstein-Institut, Max-Planck-Institut f\"ur Gravitationsphysik, D-14476 Golm, Germany}
\affiliation{\AG}
\newcommand*{\AH}{Albert-Einstein-Institut, Max-Planck-Institut f\"ur Gravitationsphysik, D-30167 Hannover, Germany}
\affiliation{\AH}
\newcommand*{\AN}{Australian National University, Canberra, 0200, Australia}
\affiliation{\AN}
\newcommand*{\CH}{California Institute of Technology, Pasadena, CA  91125, USA}
\affiliation{\CH}
\newcommand*{\DO}{California State University Dominguez Hills, Carson, CA  90747, USA}
\affiliation{\DO}
\newcommand*{\CA}{Caltech-CaRT, Pasadena, CA  91125, USA}
\affiliation{\CA}
\newcommand*{\CU}{Cardiff University, Cardiff, CF2 3YB, United Kingdom}
\affiliation{\CU}
\newcommand*{\CL}{Carleton College, Northfield, MN  55057, USA}
\affiliation{\CL}
\newcommand*{\CO}{Columbia University, New York, NY  10027, USA}
\affiliation{\CO}
\newcommand*{\TDAMSUT}{Department of Advanced Materials Science, The University of Tokyo,  Kashiwa, Chiba 277-8561, Japan} 
\affiliation{\TDAMSUT}
\newcommand*{\TDAUT}{Department of Astronomy, The University of Tokyo,  Bunkyo-ku, Tokyo 113-0033, Japan} 
\affiliation{\TDAUT}
\newcommand*{\TDPHU}{Department of Physics, Hiroshima University,  Higashi-Hiroshima, Hiroshima 739-8526, Japan} 
\affiliation{\TDPHU}
\newcommand*{\TDPMUE}{Department of Physics, Miyagi University of Education,  Aoba Aramaki, Sendai 980-0845, Japan} 
\affiliation{\TDPMUE}
\newcommand*{\TDPUT}{Department of Physics, The University of Tokyo,  Bunkyo-ku, Tokyo 113-0033, Japan} 
\affiliation{\TDPUT}
\newcommand*{\TERIUT}{Earthquake Research Institute, The University of Tokyo,  Bunkyo-ku, Tokyo 113-0033, Japan} 
\affiliation{\TERIUT}
\newcommand*{\TFSKU}{Faculty of Science, Kyoto University,  Sakyo-ku, Kyoto 606-8502, Japan} 
\affiliation{\TFSKU}
\newcommand*{\TFSNU}{Faculty of Science, Niigata University,  Niigata, Niigata 950-2102, Japan} 
\affiliation{\TFSNU}
\newcommand*{\TFSTHU}{Faculty of Science and Technology, Hirosaki University,  Hirosaki, Aomori 036-8561, Japan} 
\affiliation{\TFSTHU}
\newcommand*{\TGSASUT}{Graduate School of Arts and Sciences, The University of Tokyo,  Meguro-ku, Tokyo 153-8902, Japan} 
\affiliation{\TGSASUT}
\newcommand*{\TGSSOCU}{Graduate School of Science, Osaka City University,  Sumiyoshi-ku, Osaka 558-8585, Japan} 
\affiliation{\TGSSOCU}
\newcommand*{\TGSSOU}{Graduate School of Science, Osaka University,  Toyonaka, Osaka 560-0043, Japan} 
\affiliation{\TGSSOU}
\newcommand*{\TGSSTU}{Graduate School of Science, Tohoku University,  Sendai, Miyagi 980-8578, Japan} 
\affiliation{\TGSSTU}
\newcommand*{\THEARO}{High Energy Accelerator Research Organization,  Tsukuba, Ibaraki 305-0801, Japan} 
\affiliation{\THEARO}
\newcommand*{\HC}{Hobart and William Smith Colleges, Geneva, NY  14456, USA}
\affiliation{\HC}
\newcommand*{\TICRRUT}{Institute for Cosmic Ray Research, The University of Tokyo,  Kashiwa, Chiba 277-8582, Japan} 
\affiliation{\TICRRUT}
\newcommand*{\TILSUEC}{Institute for Laser Science, University of Electro-Communications,  Chofugaoka, Chofu, Tokyo 182-8585, Japan} 
\affiliation{\TILSUEC}
\newcommand*{\IU}{Inter-University Centre for Astronomy  and Astrophysics, Pune - 411007, India}
\affiliation{\IU}
\newcommand*{\CT}{LIGO - California Institute of Technology, Pasadena, CA  91125, USA}
\affiliation{\CT}
\newcommand*{\LM}{LIGO - Massachusetts Institute of Technology, Cambridge, MA 02139, USA}
\affiliation{\LM}
\newcommand*{\LO}{LIGO Hanford Observatory, Richland, WA  99352, USA}
\affiliation{\LO}
\newcommand*{\LV}{LIGO Livingston Observatory, Livingston, LA  70754, USA}
\affiliation{\LV}
\newcommand*{\LU}{Louisiana State University, Baton Rouge, LA  70803, USA}
\affiliation{\LU}
\newcommand*{\LE}{Louisiana Tech University, Ruston, LA  71272, USA}
\affiliation{\LE}
\newcommand*{\LL}{Loyola University, New Orleans, LA 70118, USA}
\affiliation{\LL}
\newcommand*{\MP}{Max Planck Institut f\"ur Quantenoptik, D-85748, Garching, Germany}
\affiliation{\MP}
\newcommand*{\MS}{Moscow State University, Moscow, 119992, Russia}
\affiliation{\MS}
\newcommand*{\ND}{NASA/Goddard Space Flight Center, Greenbelt, MD  20771, USA}
\affiliation{\ND}
\newcommand*{\NA}{National Astronomical Observatory of Japan, Tokyo  181-8588, Japan}
\affiliation{\NA}
\newcommand*{\TNIAIST}{National Institute of Advanced Industrial Science and Technology,  Tsukuba, Ibaraki 305-8563, Japan} 
\affiliation{\TNIAIST}
\newcommand*{\NO}{Northwestern University, Evanston, IL  60208, USA}
\affiliation{\NO}
\newcommand*{\TOU}{Ochanomizu University,  Bunkyo-ku, Tokyo 112-8610, Japan} 
\affiliation{\TOU}
\newcommand*{\TPEDTU}{Precision Engineering Division, Faculty of Engineering, Tokai University,  Hiratsuka, Kanagawa 259-1292, Japan} 
\affiliation{\TPEDTU}
\newcommand*{\TRIKEN}{RIKEN,  Wako, Saitaka 351-0198, Japan} 
\affiliation{\TRIKEN}
\newcommand*{\SC}{Salish Kootenai College, Pablo, MT  59855, USA}
\affiliation{\SC}
\newcommand*{\SE}{Southeastern Louisiana University, Hammond, LA  70402, USA}
\affiliation{\SE}
\newcommand*{\SA}{Stanford University, Stanford, CA  94305, USA}
\affiliation{\SA}
\newcommand*{\SR}{Syracuse University, Syracuse, NY  13244, USA}
\affiliation{\SR}
\newcommand*{\PU}{The Pennsylvania State University, University Park, PA  16802, USA}
\affiliation{\PU}
\newcommand*{\TC}{The University of Texas at Brownsville and Texas Southmost College, Brownsville, TX  78520, USA}
\affiliation{\TC}
\newcommand*{\TTDU}{Tokyo Denki University,  Chiyoda-ku, Tokyo 101-8457, Japan} 
\affiliation{\TTDU}
\newcommand*{\TTIT}{Tokyo Institute of Technology,  Meguro-ku, Tokyo 152-8551, Japan} 
\affiliation{\TTIT}
\newcommand*{\TR}{Trinity University, San Antonio, TX  78212, USA}
\affiliation{\TR}
\newcommand*{\HU}{Universit{\"a}t Hannover, D-30167 Hannover, Germany}
\affiliation{\HU}
\newcommand*{\BB}{Universitat de les Illes Balears, E-07122 Palma de Mallorca, Spain}
\affiliation{\BB}
\newcommand*{\BR}{University of Birmingham, Birmingham, B15 2TT, United Kingdom}
\affiliation{\BR}
\newcommand*{\FA}{University of Florida, Gainesville, FL  32611, USA}
\affiliation{\FA}
\newcommand*{\GU}{University of Glasgow, Glasgow, G12 8QQ, United Kingdom}
\affiliation{\GU}
\newcommand*{\MU}{University of Michigan, Ann Arbor, MI  48109, USA}
\affiliation{\MU}
\newcommand*{\OU}{University of Oregon, Eugene, OR  97403, USA}
\affiliation{\OU}
\newcommand*{\RO}{University of Rochester, Rochester, NY  14627, USA}
\affiliation{\RO}
\newcommand*{\UW}{University of Wisconsin-Milwaukee, Milwaukee, WI  53201, USA}
\affiliation{\UW}
\newcommand*{\VC}{Vassar College, Poughkeepsie, NY 12604}
\affiliation{\VC}
\newcommand*{\TWU}{Waseda University,  Shinjyuku-ku, Tokyo 169-8555, Japan} 
\affiliation{\TWU}
\newcommand*{\WU}{Washington State University, Pullman, WA 99164, USA}
\affiliation{\WU}
\newcommand*{\TYITPKU}{Yukawa Institute for Theoretical Physics, Kyoto University,  Sakyo-ku, Kyoto 606-8502, Japan} 
\affiliation{\TYITPKU}



\author{B.~Abbott}    \affiliation{\CT}
\author{R.~Abbott}    \affiliation{\LV}
\author{R.~Adhikari}    \affiliation{\CT}
\author{A.~Ageev}    \affiliation{\MS}  \affiliation{\SR}
\author{J.~Agresti}    \affiliation{\CT}
\author{B.~Allen}    \affiliation{\UW}
\author{J.~Allen}    \affiliation{\LM}
\author{R.~Amin}    \affiliation{\LU}
\author{S.~B.~Anderson}    \affiliation{\CT}
\author{W.~G.~Anderson}    \affiliation{\TC}
% ...

 \collaboration{The LIGO Scientific Collaboration, http://www.ligo.org}
 \noaffiliation


\author{T.~Akutsu} \affiliation{\TDAUT}
\author{M.~Ando} \affiliation{\TDPUT}
\author{K.~Arai} \affiliation{\NA}
\author{A.~Araya} \affiliation{\TERIUT}
\author{H.~Asada} \affiliation{\TFSTHU}
\author{Y.~Aso} \affiliation{\TDPUT}
\author{P.~Beyersdorf} \affiliation{\NA}
\author{Y.~Fujiki} \affiliation{\TFSNU}
\author{M.-K.~Fujimoto} \affiliation{\NA} 
\author{R.~Fujita} \affiliation{\TGSSOU}
\author{M.~Fukushima} \affiliation{\NA}
\author{T.~Futamase} \affiliation{\TGSSTU}
\author{Y.~Hamuro} \affiliation{\TFSNU}
\author{T.~Haruyama} \affiliation{\THEARO}
\author{K.~Hayama} \affiliation{\NA}
\author{H.~Iguchi} \affiliation{\TTIT}
\author{Y.~Iida} \affiliation{\TDPUT}
\author{K.~Ioka} \affiliation{\TGSSOU}
\author{H.~Ishizuka} \affiliation{\TICRRUT}
\author{N.~Kamikubota} \affiliation{\THEARO}
\author{N.~Kanda} \affiliation{\TGSSOCU}
\author{T.~Kaneyama} \affiliation{\TFSNU}
\author{Y.~Karasawa} \affiliation{\TGSSTU}
\author{K.~Kasahara} \affiliation{\TICRRUT}
\author{T.~Kasai} \affiliation{\TFSTHU}
\author{M.~Katsuki} \affiliation{\TGSSOCU}
\author{S.~Kawamura} \affiliation{\NA}
\author{M.~Kawamura} \affiliation{\TFSNU}
\author{F.~Kawazoe} \affiliation{\TOU}
\author{Y.~Kojima} \affiliation{\TDPHU}
\author{K.~Kokeyama} \affiliation{\TOU}
\author{K.~Kondo} \affiliation{\TICRRUT}
\author{Y.~Kozai} \affiliation{\NA}
\author{H.~Kudo} \affiliation{\TFSKU}
\author{K.~Kuroda} \affiliation{\TICRRUT}
\author{T.~Kuwabara} \affiliation{\TFSNU}
\author{N.~Matsuda} \affiliation{\TTDU}
\author{N.~Mio} \affiliation{\TDAMSUT}
\author{K.~Miura} \affiliation{\TDPMUE}
\author{S.~Miyama} \affiliation{\NA}
\author{S.~Miyoki} \affiliation{\TICRRUT}
\author{H.~Mizusawa} \affiliation{\TFSNU}
\author{S.~Moriwaki} \affiliation{\TDAMSUT}
\author{M.~Musha} \affiliation{\TILSUEC}
\author{Y.~Nagayama} \affiliation{\TGSSOCU}
\author{K.~Nakagawa} \affiliation{\TILSUEC}
\author{T.~Nakamura} \affiliation{\TFSKU}
\author{H.~Nakano} \affiliation{\TGSSOCU}
\author{K.~Nakao} \affiliation{\TGSSOCU} 
\author{Y.~Nishi} \affiliation{\TDPUT}
\author{K.~Numata} \affiliation{\TDPUT}
\author{Y.~Ogawa} \affiliation{\THEARO}
\author{M.~Ohashi} \affiliation{\TICRRUT}
\author{N.~Ohishi} \affiliation{\NA}
\author{A.~Okutomi} \affiliation{\TICRRUT}
\author{K.~Oohara} \affiliation{\TFSNU} 
\author{S.~Otsuka} \affiliation{\TDPUT}
\author{Y.~Saito} \affiliation{\THEARO}
\author{S.~Sakata} \affiliation{\TOU}
\author{M.~Sasaki} \affiliation{\TGSSOU}
\author{N.~Sato} \affiliation{\THEARO}
\author{S.~Sato} \affiliation{\NA}
\author{Y.~Sato} \affiliation{\TILSUEC}
\author{K.~Sato} \affiliation{\TPEDTU}
\author{A.~Sekido} \affiliation{\TWU}
\author{N.~Seto} \affiliation{\TGSSOU}
\author{M.~Shibata} \affiliation{\TGSASUT}
\author{H.~Shinkai} \affiliation{\TRIKEN}
\author{T.~Shintomi} \affiliation{\THEARO}
\author{K.~Soida} \affiliation{\TDPUT}
\author{K.~Somiya} \affiliation{\TDAMSUT}
\author{T.~Suzuki} \affiliation{\THEARO}
\author{H.~Tagoshi} \affiliation{\TGSSOU}
\author{H.~Takahashi} \affiliation{\TFSNU}
\author{R.~Takahashi} \affiliation{\NA}
\author{A.~Takamori} \affiliation{\TDPUT}
\author{S.~Takemoto} \affiliation{\TFSKU}
\author{K.~Takeno} \affiliation{\TDAMSUT}
\author{T.~Tanaka} \affiliation{\TYITPKU}
\author{K.~Taniguchi} \affiliation{\TGSASUT}
\author{T.~Tanji} \affiliation{\TDAMSUT}
\author{D.~Tatsumi} \affiliation{\NA}
\author{S.~Telada} \affiliation{\TNIAIST}
\author{M.~Tokunari} \affiliation{\TICRRUT}
\author{T.~Tomaru} \affiliation{\THEARO}
\author{K.~Tsubono} \affiliation{\TDPUT}
\author{N.~Tsuda} \affiliation{\TPEDTU}
\author{Y.~Tsunesada} \affiliation{\NA}
\author{T.~Uchiyama} \affiliation{\TICRRUT}
\author{K.~Ueda} \affiliation{\TILSUEC} 
\author{A.~Ueda} \affiliation{\NA}
\author{K.~Waseda} \affiliation{\NA}
\author{A.~Yamamoto} \affiliation{\THEARO}
\author{K.~Yamamoto} \affiliation{\TICRRUT}
\author{T.~Yamazaki} \affiliation{\NA}
\author{Y.~Yanagi} \affiliation{\TOU}
\author{J.~Yokoyama} \affiliation{\TGSSOU}
\author{T.~Yoshida} \affiliation{\TGSSTU}
\author{Z.-H.~Zhu} \affiliation{\NA} 

 \collaboration{The TAMA Collaboration}
 \noaffiliation

%% file: P040050-06.bbl
\begin{thebibliography}{99}
\bibitem{cite:LPEF_params}
The prefiltering for TFClusters was done in two stages.  
First, a sixth order modified butterworth high pass filter 
with a cutoff frequency of 100 Hz was applied.  
Then a detector-dependent FIR linear predictor error filter 
was applied.  The LPEF filters have approximately unity rms
output, a training time of 4 seconds (from the middle of the 
lock segment being analyzed), and order of 1024.
\bibitem{foot:trigdur}
In practice we find that
the typical trigger durations are $\sim$1 ms for LIGO and $\sim$5-8 ms
for TAMA.  These are much less than the light travel times, so that the
dominant contribution to the background comes from the window term $w$.
\bibitem{foot:obstime}
The smallest nonzero 
measurable rate is slightly higher than $(46T)^{-1}$ because 
triggers within 115 s of the edge of a data segment are dropped 
in some time shifts.  The same holds for the 3-site time shifts.
\bibitem{foot:H1H2window}
We could have imposed a tighter
coincidence window between H1 and H2, since they are co-located, and
further reduced the background rate.  This is not necessary, however,
since the false rate is already too small to be measured accurately; 
see Table~\ref{tab:results}.
\end{thebibliography}
